\documentclass[reprint,groupedaddress,amsmath,amssymb,aps,prd]{revtex4-2}

\usepackage{graphicx}
\usepackage{dcolumn}
\usepackage{bm}
\usepackage{array}

\begin{document}

\title{Detection of gravitational waves in circular particle accelerators}

\author{Suvrat Rao}
 \email{suvrat.rao@uni-hamburg.de}
\author{Marcus Br\"uggen}
\author{Jochen Liske}
\affiliation{Hamburger Sternwarte, University of Hamburg, Gojenbergsweg 112, 21029 Hamburg, Germany}https://www.overleaf.com/project/5ee799654a6ac50001c62231

\date{\today}

\begin{abstract}
 Here we calculate the effects of astrophysical gravitational waves (GWs) on the travel times of proton bunch test masses in circular particle accelerators. We show that a high-precision proton bunch time-tagging detector could turn a circular particle accelerator facility into a GW observatory sensitive to millihertz (mHz) GWs. We comment on sources of noise and the technological feasibility of ultrafast single photon detectors by conducting a case study of the Large Hadron Collider (LHC) at CERN. 
\end{abstract}

\maketitle

\section{Introduction}

In 2016, the first direct detection of gravitational waves (GWs) from the merger of two black holes was reported by LIGO \cite{10.1103/PhysRevLett.116.061102}. LIGO \cite{Abbott2009} uses laser interferometry to measure the GW-induced changes in the path length (equivalently, travel time) of light signals. While Earth-based interferometers such as LIGO are sensitive to high-frequency gravitational waves from compact binary mergers, the planned space-based interferometer LISA \cite{2017arXiv170200786A}, to be launched in 2034, will be sensitive to millihertz (mHz) GWs \cite{2019arXiv190706482B}. Millihertz GWs are produced by the inspiral, merger and quasi-normal modes of massive black hole binaries, and of extreme and intermediate mass ratio binaries (massive black hole - lighter compact object). They are also produced by white dwarf binaries within our Galaxy, which may be detected as individually resolved sources or as an unresolved stochastic background \cite{Sathyaprakash2009}. 

In laser interferometry, the light signals in the interferometer arms carry information about the respective test mass geodesics, making it one possible technique of comparing two or more geodesics, which is the fundamental requirement for detecting any gravitational phenomenon \cite{Sathyaprakash2009}. However, this technique imposes certain design limitations for detecting mHz GWs: first, the test mass mirrors at the ends of the interferometers are subject to gravity gradient (GG) noise (tiny tidal gravitational forces caused by perturbations of the local Newtonian gravitational potential due to human activity, seismic and atmospheric activities, etc.) which increases steeply below 1 Hz \cite{Sathyaprakash2009}. Second, the path length of the laser beams determines the frequency of the GW to which the interferometer is sensitive. The laser path length required for detecting mHz GWs is too large to be realized on Earth, even with multiple reflections of the laser beam between mirrors to fold the path length into a smaller physical space, as this increases the thermal noise \cite{Sathyaprakash2009}. Such path lengths can only be realized with space-based laser interferometers that are relatively free from GG noise.

However, both of the above limitations can be overcome by considering a circular particle accelerator, such as the Large Hadron Collider (LHC) at CERN \cite{Evans2008}. Here, the proton bunches act as test masses that are freely falling with a constant speed along the instantaneous tangent to the circular trajectory, and whose circular motion is affected by GWs, which leads to a periodic change in their travel time from its nominal value. Therefore, a fundamental difference between laser interferometry and this GW detection technique is that the former uses stationary test masses confined to rectilinear free-fall along the interferometer arms, while the latter uses moving test masses confined to circular tangential free fall in a particle accelerator.

In an accelerator, two geodesics are compared by the transfer of information via light signals that travel between two clocks which measure time differently: One clock is made up of the circulation of proton bunches, and the other can be a high-precision optical clock connected to a single photon detector. Geodesic information is transferred from the proton bunch clock to the detector clock via visible synchrotron radiation photons emitted by the proton bunches due to the magnetic field. GWs affect the proton bunch clock by changing the geodesics of the proton bunch test masses. This is detected by comparison with the optical clock, which is not affected by GWs in the same way. Here, the travel time of the light signal does not change much due to GWs because the proton bunch is very close to the detector when it transmits photons. This is different from laser interferometry where the GWs induce a change in the light travel time over long distances.

Since the proton bunches in the particle accelerator can, in principle, be confined to a circular trajectory for a very long time, their geodesics can be monitored over a sufficiently long period for mHz GWs to cause a measurable change in the proton bunch travel time.

\section{Effect of GWs on proton bunch travel times}

A GW of astrophysical origin that propagates along the $z_{\rm GW}$-axis in a reference frame $(t,x_{p},y_{p},z_{\rm GW})$, distorts spacetime in a manner given by the linearized gravity metric in the transverse-traceless gauge \cite{Pirani2009, PhysRev.105.1089},
\begin{eqnarray}
dS^2 = &&-dt^2 + \bigl(1+h_{+}(t)\bigr)dx_{p}^2 + \bigl(1-h_{+}(t)\bigr)dy_{p}^2 +\nonumber\\
&&2h_{\times}(t)dx_{p}dy_{p} + dz_{\rm GW}^2 .
\end{eqnarray}
The $+$ and $\times$ symbols denote the two possible polarizations for the time-varying GW strain. The purely temporal dependence of the GW strain is precise within a region about the origin with a length scale that is small compared to the GW wavelength, otherwise the GW strain would also have a spatial dependence of the form $h(t,z) = h(t)\,e^{i(kz-\omega t)}$.

In an observer's reference frame $(t,x,y,z)$, where the GW propagation direction has an azimuth $\phi$, inclination $\theta$, and polarization angle $\psi$, the metric is found by performing a coordinate transformation from $(t,x_{p},y_{p},z_{\rm GW})$ to $(t,x,y,z)$ via suitable elemental rotations. After doing so and switching to cylindrical coordinates $(t,r,\alpha,z)$, if we set $r=R$, $dr=0$ and $dz=0$, we see the effect of GWs along a circular arc of radius $R$. The effective metric takes the form,
\begin{equation}
   dS^2 = -dt^2 + \bigl(1+h_{\theta\phi\psi}(t,\alpha)\bigr)R^2d\alpha^{2} ,
   \label{metric}
\end{equation}
and the effective GW strain, $h_{\theta\phi\psi}(t, \alpha)$, has the complex expression,
\begin{equation}
\begin{aligned}
   h_{\theta\phi\psi}(t, \alpha) = {} & h_{+}(t)\Bigl(f_{s}^{+}\sin^2{\alpha}+f_{c}^{+}\cos^2{\alpha}+f_{sc}^{+}\sin{2\alpha}\Bigr) +\\
   &h_{\times}(t)\Bigl(f_{s}^{\times}\sin^2{\alpha}+f_{c}^{\times}\cos^2{\alpha}+f_{sc}^{\times}\sin{2\alpha}\Bigr),
\end{aligned}
\label{strain}
\end{equation}
where,
\begin{equation}
\begin{matrix}
f_{s}^{+}=(\cos^2{\theta}\cos^2{\phi}-\sin^2{\phi})\cos{2\psi}-(\cos{\theta}\sin{2\phi})\sin{2\psi},\\
f_{c}^{+}=(\cos^2{\theta}\sin^2{\phi}-\cos^2{\phi})\cos{2\psi}+(\cos{\theta}\sin{2\phi})\sin{2\psi},\\
f_{sc}^{+}=\bigl(\frac{1}{2}(1+\cos^2{\theta})\sin2{\phi}\bigr)\cos{2\psi}+(\cos{\theta}\cos{2\phi})\sin{2\psi},\\
f_{s}^{\times}=(\cos^2{\theta}\cos^2{\phi}-\sin^2{\phi})\sin{2\psi}+(\cos{\theta}\sin{2\phi})\cos{2\psi},\\
f_{c}^{\times}=(\cos^2{\theta}\sin^2{\phi}-\cos^2{\phi})\sin{2\psi}-(\cos{\theta}\sin{2\phi})\cos{2\psi},\\
f_{sc}^{\times}=\bigl(\frac{1}{2}(1+\cos^2{\theta})\sin2{\phi}\bigr)\sin{2\psi}-(\cos{\theta}\cos{2\phi})\cos{2\psi}.
\end{matrix}
\end{equation}

The travel time of test masses in a circular particle accelerator is thus calculated by integrating the timelike geodesic equations of the effective metric,
\begin{equation}
\begin{matrix}
\frac{d^2t}{d\tau^2}+\frac{1}{2}\frac{dh_{\theta\phi\psi}(t)}{dt}\bigl(\frac{dl}{d\tau}\bigr)^2 =0, \\~\\
\frac{d^2l}{d\tau^2}+\frac{1}{1+h_{\theta\phi\psi}(t)}\frac{dh_{\theta\phi\psi}(t)}{dt}\bigl(\frac{dl}{d\tau}\bigr)\bigl(\frac{dt}{d\tau}\bigr) =0, \\~\\
-\bigl(\frac{dt}{d\tau}\bigr)^2+\bigl(1+h_{\theta\phi\psi}(t)\bigr)\bigl(\frac{dl}{d\tau}\bigr)^2=g_{\mu\nu}\dot{dx^\mu}\dot{dx^\nu}=-1. \\~\\
\end{matrix}
\label{geodesic}
\end{equation}

$Rd\alpha=dl$ represents the arc length; $\frac{dt}{d\tau}=\gamma$ is the Lorentz factor; $\frac{dl}{d\tau}=\gamma v$, where $v$ is the circular velocity. Here we write the GW strain as a function of time only, with the foresight that the geodesic solution for the angular coordinate would be a function of time in a circular trajectory, $\alpha = \alpha(t)$. Solving the first and third geodesic equations gives the time-variation of the test mass velocity due to the GWs,
\begin{equation}
v^2(t) = \frac{\bigl(1+h_{\theta\phi\psi}(t_0)\bigr).v^2(t_0)}{\bigl(1+h_{\theta\phi\psi}(t_0)\bigr).v^2(t_0) + \bigl(1+h_{\theta\phi\psi}(t)\bigr).\bigl(1-v^2(t_0)\bigr)} .
\label{velocity}
\end{equation}

We note that $v(t_0)$, the velocity of the particles at the beginning of the observation run, $t_0$, differs from that of the nominal case, $v_0$, and are related as $v^2(t_0) = v_0^2\bigl(1+h_{\theta\phi\psi}(t_0)\bigr)$. This is a result of the frame dependence of velocity measurement for timelike signals. This issue does not exist for lightlike signals, which have a frame-independent, constant velocity. The travel time of the test masses is found by integrating the third geodesic equation, written as,
\begin{equation}
\bigl(1+h_{\theta\phi\psi}(t)\bigr)dl^2 = v^2(t)dt^2 .
\end{equation}

We make the substitution for $v^2(t)$ and use the binomial approximation, as $h_{\theta\phi\psi}(t)$ is small. We revert back to metric units from natural units. Thus, in terms of the constant velocity, $v_0$, of the nominal case, we get the relation,
\begin{equation}
dl = v_0dt\Biggl(1-\Bigl(1-\frac{v_0^2}{2c^2}\Bigr)\bigl(h_{\theta\phi\psi}(t)-h_{\theta\phi\psi}(t_0)\bigr)\Biggr) .
\label{integrand}
\end{equation}

In the nominal case, the travel time of a test mass with constant circular velocity $v_0$, making $n$ (not necessarily whole) turns in a circular accelerator of radius $R$, is $T= \frac{2\pi nR}{v_0} = \frac{L}{v_0}$. We integrate Eq.~(\ref{integrand}) over the nominal travel time and make the interpretation that for the same travel time, the test mass covers a slightly different path length, by $\Delta L$, in the presence of GWs. Thus, we have,
\begin{equation}
L + \Delta L = v_0T - v_0\Bigl(1-\frac{v_0^2}{2c^2}\Bigr)\int_{t_0}^{t_0+T}\bigl(h_{\theta\phi\psi}(t)-h_{\theta\phi\psi}(t_0)\bigr)dt .
\end{equation}

The change in travel time from its nominal value is therefore, the extra time that the test mass would take to cover the length $\Delta L$, with the final velocity $v(t_0 + T)$. Even so, since we neglect terms of $\mathcal{O}(h^2)$ and higher, the calculation reduces to simply $\frac{-\Delta L}{v_0}$. Thus, in the presence of GWs, the deviation of the test mass travel time from its nominal value is given by the general expression,
\begin{equation}
   \Delta T_{\rm GW} = \Bigl(1-\frac{v_{0}^2}{2c^2}\Bigr)\int_{t_{0}}^{t_{0}+T}\bigl(h_{\theta\phi\psi}(t, \alpha(t))-h_{\theta\phi\psi}(t_{0}, \alpha_0)\bigr)dt .
   \label{dtcircular}
\end{equation}

The detector is located at an initial angular coordinate $\alpha_0$, and begins observing at time $t_0$. Here, we take the angular coordinate of the circulating test mass to vary with time as in the nominal case, $\alpha(t)=\alpha_0 + \frac{v_0}{R}(t-t_0)$, since all terms of order $\mathcal{O}(h^2)$ and higher can be neglected. We may arbitrarily set $\alpha_0=0$, since the relative orientation of the GW can be compensated in the choice of the azimuth, $\phi$.

\begin{figure*}[t]
\includegraphics[width=0.5301\linewidth]{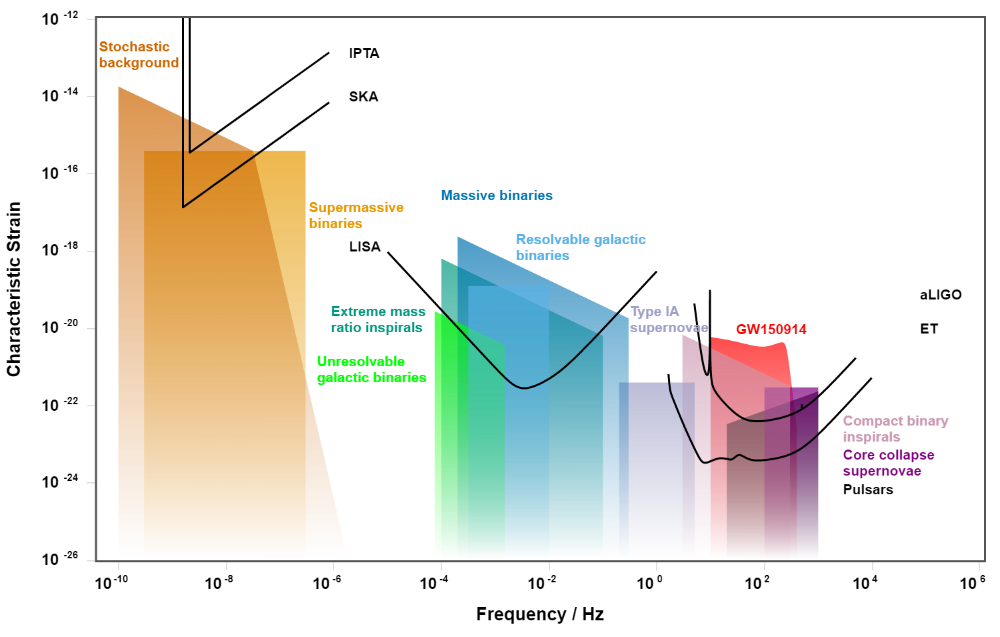}
\includegraphics[width=0.4645\linewidth]{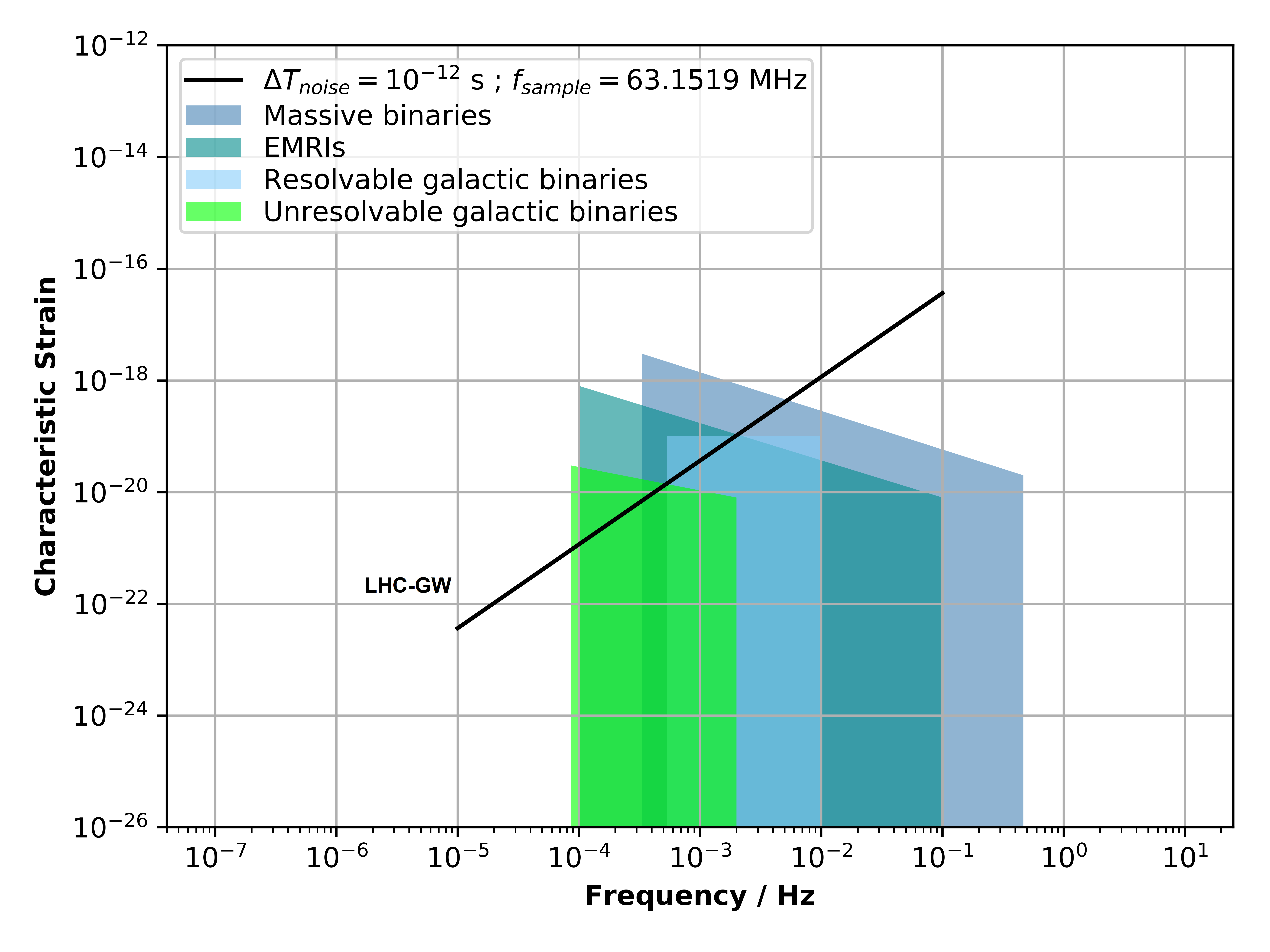}
\caption{The GW sensitivity curves of other GW detectors and the LHC (left panel adopted from \cite{Moore_2014}). With an overall noise magnitude of $\sim 10^{-12}$ s due to noise sources, and the maximum LHC sampling rate of $\approx 63.1519$ MHz, we see that it would be possible to detect the strong mHz GWs produced by extreme mass ratio inspirals, massive black hole binaries and Galactic binaries, resulting in some overlap with the LISA sensitivity curve. GWs from binary supermassive black holes are also accessible in principle, but detecting them would require continuous measurements over unfeasibly long periods ($>10^6$ s).}
\label{lhc_gw_sensitivity}
\end{figure*}

We obtain an explicit expression for Eq.~(\ref{dtcircular}) by making the ansatz, for simplicity, of a linearly polarized sinusoidal GW (see Appendix~\ref{app_A} for detailed calculations). Using this assumption, we find that the order of magnitude of the travel time variation over a GW period, $1/f_{\rm GW}$, is
\begin{equation}
   \Delta T_{\rm GW} \sim a\frac{h_0}{f_{\rm GW}} ,
   \label{dtmagnitude}
\end{equation}
where $a$ is a constant of order $\sim 10^{-2}$. Strong mHz GWs with characteristic strain and frequency of order $h_0\sim10^{-18}$, $f_{\rm GW}\sim 10^{-4}$ Hz, respectively, would cause, at best, $\Delta T_{\rm GW}\sim10^{-16}$ s, which corresponds to a shift in test mass positions by $\sim10^{-8}$ m.

\section{GW Sensitivity Curve}

Separate timing detectors would need to time-tag the proton bunches constituting the counter-rotating, ultrarelativistic proton beams in the two LHC vacuum tubes, which can be made to circulate without collisions. The LHC utilizes superconducting electromagnets along with beam collimators to impose a highly precise circular trajectory of the proton bunches in the focused proton beam. Further, the LHC Radiofrequency (RF) system \cite{10.1103/PhysRevSTAB.13.102801} keeps the proton bunches circulating at a constant speed and also maintains the bunch length ($\approx 30$ cm) of the 2808 proton bunches separated by $\approx 25$ ns travel time. In the presence of strong mHz GWs, the travel time of all proton bunches will periodically deviate from its nominal value of $\approx n \times 89\, \mu$s ($n$ being the fractional number of revolutions and $89\, \mu$s the revolution period) in a continuous manner by up to $\Delta T_{\rm GW}\sim10^{-16}$ s, with a frequency of $f_{\rm GW}$. Every timing measurement is, therefore, a sample of the continuous, time-varying GW signal.

Each LHC vacuum tube can contain 2808 proton bunches, circulating at $\approx 11,245$ Hz. Hence, the maximum sampling rate (i.e.\ the rate at which the detectors time-tag the proton bunches) at the LHC is $\approx 2\times2808\times11245=63.1519$ MHz. A high sampling rate improves signal detection by improving noise filtering techniques such as matched filtering \cite{Sathyaprakash2009}, thus potentially enabling the detection of the GW signal even if it is buried within the noise caused by various error sources which perturb the proton bunch travel time or its measurement. To detect GWs, the travel time difference magnitude due to GWs given by Eq.~(\ref{dtmagnitude}), must be greater than the overall noise magnitude, $\Delta T_{\rm noise}$. However, employing matched filtering, the effective noise is reduced by the square root of the number of data-points taken within the observing time i.e. the sampling rate, $f_{\rm sample}$, times the observing time, $T_{obs}$. Hence, we have
\begin{equation}
   \Delta T_{\rm GW} \gtrsim \frac{\Delta T_{\rm noise}}{\sqrt{f_{\rm sample}\,T_{\rm obs}}} .
   \label{sensitivity}
\end{equation}

We assume that the significant portion of the GW signal lasts for a duration of 10 times its period, i.e.  $T_{\rm obs} \sim 10/f_{\rm GW}$. Inserting this in the above inequality, and the time-difference magnitude estimate Eq.~(\ref{dtmagnitude}) for a circular accelerator GW observatory, we obtain the LHC-GW sensitivity curve, shown in Fig.~\ref{lhc_gw_sensitivity},
\begin{equation}
   {h_{0} \gtrsim \frac{b\,\Delta T_{\rm noise}}{\sqrt{f_{\rm sample}}}(f_{\rm GW})^{3/2} } ,
\end{equation} 
where $b$ is a constant of order 10. 

\section{Noise sources}

Below, we estimate the magnitudes of various sources of noise in terms of the nominal proton bunch travel time, $T$:

\subsection{Quantum noise}

The  uncertainty principle sets a lower limit on the proton bunch timing uncertainty. The uncertainty in the relativistic momentum of the test mass, in terms of the uncertainty in its velocity, is given by,
\begin{equation}
    \Delta p = m_0\Delta(\gamma v)\Bigl|_{v=v_0} = m_0\Bigl(1-\frac{v_0^2}{c^2}\Bigr)^{-3/2}\Delta v .
\end{equation}
Therefore, the uncertainty relation reads,
\begin{equation}
    \Delta L \sim \frac{\hslash}{2}.\frac{1}{m_0\Delta v}\Bigl(1-\frac{v_0^2}{c^2}\Bigr)^{3/2} .
\end{equation}
Since the travel time is $T=L/v_0$, the uncertainty in timing due to uncertainties in position and velocity measurements is given by,
\begin{equation}
    \Delta T = \frac{\Delta L}{v_0} + \frac{L\Delta v}{v_0^2} .
\end{equation}
Substituting $\Delta L$ and $L=v_0T$, we get,
\begin{equation}
    \Delta T \sim \frac{\hslash}{2m_0v_0}\Bigl(1-\frac{v_0^2}{c^2}\Bigr)^{3/2}\frac{1}{\Delta v} + \frac{T}{v_0}\Delta v .
\end{equation}
The timing uncertainty can be minimized for a suitable uncertainty of velocity measurement, giving,
\begin{equation}
\Delta T \sim \sqrt{\frac{2\hslash T}{m_0}}\frac{1}{v_0}\Bigl(1-\frac{v_0^2}{c^2}\Bigr)^{3/4} .
\end{equation}
In the LHC, protons circulate with a period of $\approx 89\,\mu$s at a speed which is approximately $3$ m/s slower than the speed of light. Assuming that a detection takes place every revolution, then the quantum noise is,
\begin{equation}
\Delta T_{\rm quantum} \sim 10^{-20} s .
\end{equation}

\subsection{Gravity gradient (GG) noise}

GG noise, unlike GWs, is a tidal gravitational effect. For interferometric GW detectors, GG noise mimics the effect of a stochastic background of GWs, and is dominant at low frequencies \cite{PhysRevD.58.122002, PhysRevD.60.082001}. In the LHC, ultra-relativistic proton bunches circulate with a frequency of $\approx 11,245$ Hz, at an underground depth of 50-175 m. The travel time change caused by tidal gravitational forces due to terrestrial and non-terrestrial masses acting directly on the proton bunches would be greatly attenuated by their high frequency circulation. To show this, we make a simple estimate using a classical treatment, by considering the tidal gravitational acceleration on the proton bunches due to a mass M, at a distance R. For terrestrial masses close to the LHC, we find,
\begin{equation}
    \Delta T_{\rm GG} \sim \Bigl(\frac{GM}{2 h v_0^2}\Bigr)T ,
\end{equation}
where $h$ is the depth of the LHC below the mass, and $v_0$ is the circular speed of the proton bunches, close to the speed of light. For a typical 250,000 ton skyscraper of a few hundred meters height, the GG noise over a duration of $T=10^4$ s is $\Delta T_{\rm GG} \sim 10^{-18}$ s, but in reality this value would be roughly averaged out due to all the masses surrounding the LHC, making the GG noise due to nearby terrestrial objects negligible compared to the effect of mHz GWs. On the other hand, for masses far away from the LHC, we find,
\begin{equation}
    \Delta T_{\rm GG} \sim \Bigl(\frac{2GM L^2}{R^3 \pi v_0^2}\Bigr)T ,
\end{equation}
where $L$ is the circumference of the LHC ($\approx 26.7$ km). Over a duration of $T=10^4$ s, the tides due to the Sun and Moon (even if assumed to be aligned during New Moon and having a fixed, edge-on orientation with the LHC ring), would cause $\Delta T_{GG} \sim 10^{-18}$ s. Actually, this value would be even smaller due to a changing orientation caused by Earth's rotation and misalignment between the Sun and the Moon. However, the tides also cause a slight deformation of the LHC tunnel (by $\approx 1$ mm), over the period of a day \cite{Arnaudon1995}. A radial feedback loop allows the RF system to continuously correct the proton beam orbit, keeping it centered \cite{10.1103/PhysRevAccelBeams.20.081003, Baudrenghien:971742}. The long-term proton bunch travel time variation due to this effect can therefore be fully predicted and accounted for. Finally, we find that the Alps mountains located closer than a few hundred kilometers from LHC can also cause a significant GG noise. For the Mont Blanc, which is roughly 100 km away from the LHC, assuming a typical mass of $10^{14}$ kg, we estimate $\Delta T_{\rm GG} \sim 10^{-16}$ s over a duration of $T=10^4$ s. However, the local gravity gradient across the LHC can be mapped in detail to completely account for the net GG noise due to all stationary terrestrial mass distributions.

\subsection{Seismic noise}

Incoherent ambient seismic activity is the major source of mechanical vibrations that can cause relative motion between the LHC dipole magnets, leading to small deformations (by $\sim \mu$m) of the proton beam orbit \cite{Gamba:2648557, Steinhagen:914080, Vos:359254, Vos:274506}. However, mechanical vibrations cannot affect the longitudinal dynamics of the proton bunches since they couple only via the magnetic field. We make a simple estimate of seismic noise on the proton bunch travel times: If laid out in a straight line, the deformed orbit can be modelled as a sine wave about the ideal orbit. The deviation of the deformed orbit from the ideal orbit at a position $x$ and time $t$ is given by
\begin{equation}
    y = y_0\sin{(2\pi f_s t)}\sin{\Bigl(\frac{n\pi x}{L}\Bigr)} ,
\end{equation}
where $L$ is the circumference of the ideal orbit ($\approx 26.7$ km for LHC); $f_s$ is the frequency of seismic waves; $n$ is the harmonic number (number of antinodes of the deformed orbit with respect to the ideal orbit when laid out in a straight line over one revolution). The small deviation of travel time is proportional to the difference in path lengths of the deformed and ideal orbits,
\begin{equation}
    v_0\Delta T = \frac{1}{2}\int_{0}^{T}\Bigl( \frac{dy}{dx}\Bigr)^2\Bigl|_{x=v_0t}v_0dt .
\end{equation}
For low-frequency seismic waves, we find the estimate,
\begin{equation}
    \Delta T = \frac{n^2 \pi^2 y_0^2}{8L^2}\Bigl(T - \frac{\sin{(4\pi f_s T)}}{4\pi f_s}\Bigr)  .
\end{equation}
For mHz seismic waves causing an orbit deformation of $y_0\sim\mu$m with low harmonic number, and a duration of $T=10^4$ s, the seismic noise is,
\begin{equation}
    \Delta T_{\rm seismic} \sim 10^{-17} s .
\end{equation}
An orbit correction mechanism such as the radial feedback loop of the RF system, which can reduce the orbit deformation amplitude, will further reduce the seismic noise on the proton bunches. Lastly, the detector itself will also be directly affected by seismic vibrations, causing a significant seismic noise. To overcome this issue, a technological solution may be employed similar to the VIRGO interferometer, whose Superattenuators can attenuate seismic vibrations by more than 10 orders of magnitude \cite{Accadia2011}.

\subsection{Radiofrequency phase noise}

The precision of the RF system implementation (responsible for controlling the constant speed of the proton bunches, the bunch length, the separation between the bunches and their arrival in-phase with the oscillating electric field) leads to a jitter of the proton bunch travel times, causing a noise with a magnitude of $\Delta T_{\rm RF}\sim 10^{-12}$ s on the timescale of the measurement \cite{Baron2012}. A correction for bunch elongation over time of around 8 ps/hr in the stable beam \cite{Baron2012} must also be taken into account in the models.

\subsection{Detector noise}

The Longitudinal Density Monitor (LDM) \cite{PhysRevSTAB.15.032803} is an existing detector at the LHC, suitable for this GW detection technique. It is a single photon counting system with a timing jitter of $\Delta T_{detector}=50$ ps and dead time (duration of instrument unresponsiveness between two consecutive detections) of 77 ns that detects the visible synchrotron radiation emitted by the proton bunches to measure their longitudinal profile.  

\subsection{Photon shot noise}

The inverse of the average photon arrival rate at the detector is the additional jitter on the proton bunch time-tagging, as the arrival time of a photon at the detector can roughly vary by this value for every proton bunch. The synchrotron radiation emission pattern for ultrarelativistic protons is sharply collimated forward along the velocity of the proton, with an opening angle of $1/\gamma \sim 10^{-3}$ degrees. For a proton bunch of $\approx 30$ cm traversing the $L\approx26.7$ km LHC circumference, the emission patterns of the constituent $\sim 10^{11}$ protons significantly overlap. From the Li\'enard-Wiechert formulation of synchrotron radiation due to a bending magnetic field, the average power received by the detector over all frequencies and all solid angle as the synchrotron radiation from a proton bunch sweeps across the detector, is
\begin{equation}
    \langle P \rangle_{\rm sync} \sim 10^{11} \times \frac{2\pi e^2 c \beta^4 \gamma^4}{3 \epsilon_0 L^2} \sim 10^{-1} W.
\end{equation}
Where $\beta=v_0/c$ and $\gamma$ is the Lorentz factor. Considering the frequency distribution of the radiated energy, we obtain the equivalent frequency, $\nu_{eq}$, which gives for the total energy radiated, the same total number of photons: 
\begin{equation}
    \nu_{eq} \sim \frac{\gamma^3 c}{L} \sim 10^{16} Hz.
\end{equation}
Thus, the inverse of the average arrival rate of photons at the detector, which is the photon shot noise, is estimated to be
\begin{equation}
    \Delta T_{\rm photon} \sim \biggl(\frac{\langle P \rangle_{\rm sync}}{h\nu_{eq}}\biggr)^{-1} \sim 10^{-17} s .
\end{equation}
Since the above calculation is over all frequencies, we interpret that increasing the bandwidth of the detector can reduce the true noise below the required threshold. Moreover, the time-tagging would be less accurate near the edges of the emission pattern, where the power per solid angle is much less. 

\section{Discussions}

Presently, the significant noise at LHC is due to the detector timing jitter of 50 ps, and the sampling rate is limited by the detector dead time of 77 ns, which is longer than the bunch length ($\approx 1$ ns) and the separation between proton bunches ($\approx 25$ ns). But already, new technology such as superconducting nanowire single-photon detectors (SNSPDs) offer a low timing jitter of a few ps \cite{Korzh2020}, high photon count rate (inverse of the dead time) of a few GHz \cite{8627992}, low dark count rate (rate of recording false counts) of $10^{-3}$ Hz \cite{Shibata2015} and high detection efficiency (probability of detecting a photon which arrives at the detector) of around 98\% \cite{Reddy2019}, while also showing the potential for realizing a single device with all these merits \cite{EsmaeilZadeh2017}. Moreover, SNSPDs can function over a broad wavelength range \cite{Li2019} including  the visible-UV range, even in the presence of strong magnetic fields \cite{Polakovic2020}. A high-efficiency, broadband detection system with $\approx 1$ ps timing jitter and $\approx 1$ GHz count rate which can detect a few photons per proton bunch, would allow the accurate time-tagging of every LHC proton bunch consecutively and continuously, thus reaching the sampling rate of $\approx63.1519/2$ MHz per vacuum tube. Importantly, the detection system must have a suitable timing precision (least count) that is smaller by a few orders of magnitude than the expected peak signal strength due to GWs, in order to detect the whole waveform. The latest optical clocks offer a timing precision of $\sim 10^{-19}$ s \cite{PhysRevLett.120.103201}.

We observe that in this detection technique, certain noise sources such as gravity gradient noise and seismic noise always act as ``drag" i.e. which tend to continually build up a time delay in the proton bunch time tagging while causing no significant timing jitter. This is in contrast to laser interferometry, where these same noise sources tend instead to mimic a stochastic GW background. Meanwhile, the other noise sources in this detection technique cause only a timing jitter.

In the LHC, the RF system is present only at certain points of the circular accelerator ring. The ideal RF system causes no acceleration to be experienced by those particles which travel through it with the correct speed and timing (synchronous particles), while all other particles in the proton bunch perform longitudinal oscillations about the synchronous particle while traveling (synchrotron oscillations), because the RF system causes all non-synchronous particles to experience a suitable acceleration that tends to restore the synchronous parameters. Since an LHC proton bunch consists of $\sim 10^{11}$ particles, let us assume a continuous velocity distribution of the particles having a tiny spread around the mean (which is the synchronous velocity), and also a continuous position distribution with a spread equal to the bunch length. Considering this picture, we see that in spite of the presence of an RF system, the effect of GWs (which would equally affect all particles in the proton bunch) would be to cause continuous, small and slow (relative to the synchrotron oscillations) periodic shifts of the particle velocities and positions, as given by Eq.~(\ref{velocity}) and Eq.~(\ref{dtcircular}) respectively. Hence, we can expect the GW signal to be present in the travel time information of the proton bunches in a circular particle accelerator.

For determining the location of an astrophysical GW source, triangulation via the time delay between three or more GW observatories at different locations detecting the same event is needed, as done by the extended LIGO/VIRGO collaboration \cite{Abbott2018}. The space-based LISA detectors are expected to have a sky-localization of 1-100 deg$^2$, which can improve in a network with other space-based detectors \cite{Ruan2020}. The LISA orbit would lag the Earth orbit by $\approx 20$ degrees, implying an optimistic improvement by roughly 1.5 orders of magnitude of the angular resolution, for massive binary GW events detected by both LISA and LHC. Thus, the network of LISA with the potential LHC-GW observatory can achieve a significantly higher precision of localizing the GW source in the sky.

\begin{figure}[t]
\includegraphics[width=\linewidth]{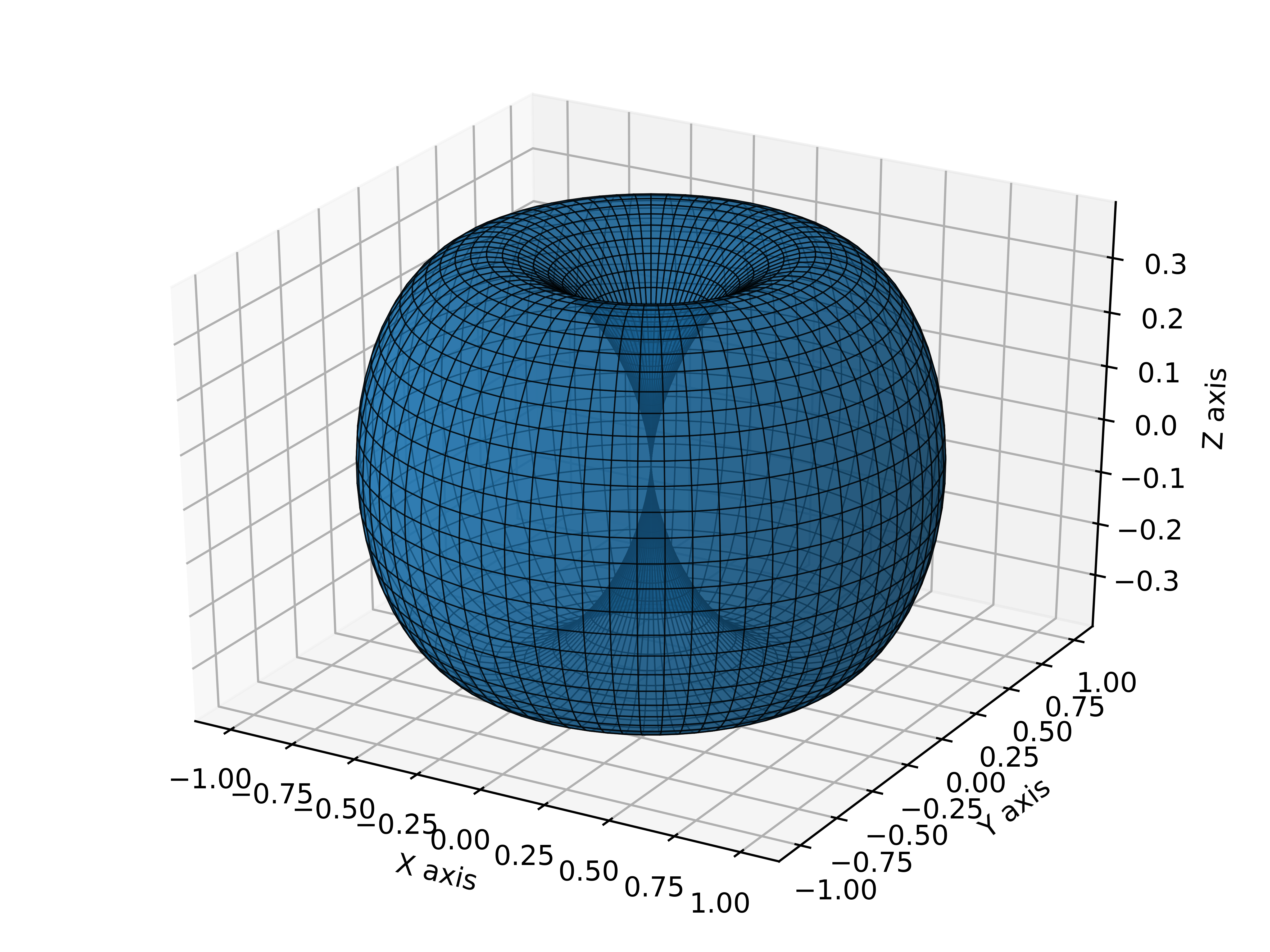}
\caption{The LHC-GW antenna pattern. Shown here is the approximate ‘pointing’ of a circular particle accelerator GW observatory, rms-averaged over all possible GW polarization angles (see Appendix~\ref{app_A}). The origin lies at the center of the circular accelerator with the z-axis being perpendicular to its plane.}
\label{lhc_gw_pointing}
\end{figure}

While probing mHz GWs with any Earth-based detector, the orientation of the GW with respect to the detector would vary significantly (by more than 10 degrees) over the detection period due to the Earth's rotation, and therefore, this must be taken into account in the models (see Appendix~\ref{app_B}). This may change the response of the detector if the GW orientation becomes aligned differently with the antenna pattern, shown in Fig.~\ref{lhc_gw_pointing}.

\section{Conclusion}

We conclude that a high-frequency, low timing jitter and high-precision proton bunch time-tagging detector could turn a circular particle accelerator facility into a GW observatory sensitive to mHz GWs. 

\begin{acknowledgments}
We acknowledge Valerie Domcke, Aravindh Swaminathan and Roman Schnabel for fruitful discussions. This research was supported by the Deutsche Forschungsgemeinschaft (DFG, German Research Foundation) under Germany’s Excellence Strategy – EXC 2121 Quantum Universe – 390833306.
\end{acknowledgments}

\appendix
\section{Sinusoidal GW ansatz, time-difference magnitude and antenna pattern} \label{app_A}

In Eq.~(\ref{strain}), we ignore the cross polarization term (effectively assuming a linearly polarized GW), and use a sinusoidal ansatz of the time-varying GW strain,
\begin{equation}
h_{+}(t) = h_0\sin{(2\pi f_{GW}t)} .    
\end{equation}

We substitute Eq.~(\ref{strain}) in the general expression for the travel time change,
\begin{equation}
   \Delta T_{\rm GW} = \Bigl(1-\frac{v_{0}^2}{2c^2}\Bigr)\int_{t_{0}}^{t_{0}+T}\bigl(h_{\theta\phi\psi}(t, \alpha(t))-h_{\theta\phi\psi}(t_{0}, \alpha_0)\bigr)dt .
   \label{dtcirc}
\end{equation}

In the nominal case, the travel-time of particles with constant circular velocity $v_0$, making $n$ (not necessarily whole) turns in a circular accelerator of radius $R$ is $T= \frac{2\pi nR}{v_0}$. The observing run begins at time $t_0$, from an angular coordinate, $\alpha_0$, of the detector's location. We may arbitrarily set $\alpha_0=0$, since the relative orientation of the GW can be compensated in the choice of the azimuth, $\phi$. This fixes the $x$-axis to pass through the detector location. And, we take $\alpha(t)=\frac{v_0}{R}(t-t_0)$, since all terms of order $\mathcal{O}(h^2)$ and higher can be neglected. Therefore, we have,
\begin{eqnarray}
   \Delta T_{GW} = &&h_0\Bigl(1-\frac{v_{0}^2}{2c^2}\Bigr) \nonumber\\ 
   &&\Biggl(f_{s}^{+}\int_{t_{0}}^{t_{0}+T}\sin{(2\pi f_{GW}t)}\sin^2{\Bigl(\frac{v_0}{R}(t-t_0)\Bigr)}dt \nonumber\\
   &&+f_{c}^{+}\int_{t_{0}}^{t_{0}+T}\sin{(2\pi f_{GW}t)}\cos^2{\Bigl(\frac{v_0}{R}(t-t_0)\Bigr)}dt \nonumber\\
   &&+f_{sc}^{+}\int_{t_{0}}^{t_{0}+T}\sin{(2\pi f_{GW}t)}\sin{\Bigl(2\frac{v_0}{R}(t-t_0)\Bigr)}dt \nonumber\\
   &&- f_{c}^{+}T\sin{(2\pi f_{GW}t_0)}\Biggr). \nonumber\\ 
\end{eqnarray}

Solving the time integral, and using $T=\frac{2\pi nR}{v_0}$, we get,
\begin{eqnarray}
   \Delta T_{GW} = &&h_0\Bigl(1-\frac{v_{0}^2}{2c^2}\Bigr)\times \nonumber\\
   &&\Biggl(\frac{(f_{s}^{+}+f_{c}^{+})}{2\pi f_{GW}}\sin{(\pi f_{GW}T+2\pi f_{GW}t_0)}\times\nonumber\\
   &&\sin{(\pi f_{GW}T)} - f_{c}^{+}T\sin{(2\pi f_{GW}t_0)} \nonumber\\ &&+\frac{(f_{s}^{+}-f_{c}^{+})}{4\pi}.\frac{T}{f_{GW}T+2n}\sin{(\pi f_{GW}T +2\pi n)} \times \nonumber\\
   &&\sin{(\pi f_{GW}T+2\pi n+2\pi f_{GW}t_0)} \nonumber\\
   &&+\frac{(f_{s}^{+}-f_{c}^{+})}{4\pi}.\frac{T}{f_{GW}T-2n}\sin{(\pi f_{GW}T -2\pi n)}\times \nonumber\\
   &&\sin{(\pi f_{GW}T-2\pi n+2\pi f_{GW}t_0)} \nonumber\\
   &&+\frac{f_{sc}^{+}}{4\pi}.\frac{T}{f_{GW}T-2n}\cos{(\pi f_{GW}T -2\pi n)}\times \nonumber\\
   &&\sin{(\pi f_{GW}T-2\pi n+2\pi f_{GW}t_0)} \nonumber\\
   &&-\frac{f_{sc}^{+}}{4\pi}.\frac{T}{f_{GW}T+2n}\cos{(\pi f_{GW}T +2\pi n)}\times \nonumber\\ 
   &&\sin{(\pi f_{GW}T+2\pi n+2\pi f_{GW}t_0)}\Biggr) .
\end{eqnarray}

If $v_0$ is large, close to light-speed (as is usually achieved in particle accelerators such as the LHC at CERN), implying that $n$ is also large, then the last four terms in the above result vanish, and the coefficient outside becomes $1/2$, leaving us with the result,
\begin{eqnarray}
   \Delta T_{GW} = &&\frac{1}{2}\frac{h_{0}}{f_{GW}}\Biggl( \frac{(f_{s}^{+}+f_{c}^{+})}{2\pi}.\sin{(\pi f_{GW}T+2\pi f_{GW}t_0)}\times \nonumber\\ 
   &&\sin{(\pi f_{GW}T)}- f_{c}^{+}.f_{GW}T.\sin{(2\pi f_{GW}t_0) \Biggr)} .
\end{eqnarray}

The phase term, $2\pi f_{GW}t_0$, signifies the effect on the travel time change, of the initial phase of the periodic GW signal at the beginning of the observing run, $t_0$, and here, $t_0=0$ corresponds to phase $0$. Since it is highly likely that there are no GWs at the beginning of an observing run, the condition $t_0=0$ applies, giving the final result, 
\begin{equation}
   \Delta T_{GW} = \frac{1}{4\pi}\frac{h_{0}}{f_{GW}}. F_{+} .\sin^{2}{(\pi f_{GW}T)} ,
   \label{main}
\end{equation}
\begin{equation}
   F_{+} = \sin^{2}{\theta}\cos{2\psi} .
\end{equation}

Taking an rms-average over angular parameters ($\theta, \psi$) of Eq.~(\ref{main}), we find the order of magnitude of travel time change due to GWs, of test mass particles in a circular accelerator, over the GW  period $1/f_{GW}$, to be,
\begin{equation}
   \Delta T_{GW} \sim \frac{\sqrt{3}}{16\pi}\frac{h_0}{f_{GW}} .
\end{equation}

The antenna-pattern of this GW detector, rms-averaged over all GW polarization angles, is given by
\begin{equation}
   \overline{F_{+}} = \Bigl(\frac{1}{2\pi}\int_{0}^{2\pi}\sin^{4}{\theta}\cos^{2}{2\psi}.d\psi\Bigr)^{1/2} \sim \sin^{2}{\theta} .
\end{equation}

\section{Effect of Earth's rotation} \label{app_B}
The change in travel time given by Eq.~(\ref{dtcirc}) must be recalculated by considering the time-varying GW orientation due to Earth's rotation, $\theta(t), \phi(t), \psi(t)$, the terms corresponding to which would now be inside the time integral, which must be evaluated numerically.

The following relation establishes the connection between the time-varying GW orientation $\theta(t), \phi(t), \psi(t)$, as measured from a moving coordinate system on Earth with its origin at the center of the circular accelerator, $(x,y,z)$, and the equatorial celestial coordinate system, in which the orientation of the GW is fixed:-
\begin{multline}
\begin{pmatrix}
\cos{\psi} & -\sin{\psi} & 0\\
\sin{\psi} & \cos{\psi} & 0\\
0 & 0 & 1
\end{pmatrix}\begin{pmatrix}
\cos{\theta} & 0 & \sin{\theta} \\
0 & 1 & 0\\
-\sin{\theta} & 0 & \cos{\theta}
\end{pmatrix}\begin{pmatrix}
\cos{\phi} & -\sin{\phi} & 0\\
\sin{\phi} & \cos{\phi} & 0\\
0 & 0 & 1
\end{pmatrix} = \\
\begin{pmatrix}
\cos{\psi_{eq}} & -\sin{\psi_{eq}} & 0\\
\sin{\psi_{eq}} & \cos{\psi_{eq}} & 0\\
0 & 0 & 1
\end{pmatrix}\begin{pmatrix}
\cos{\bigl(\frac{\pi}{2}-\delta_{GW}\bigr)} & 0 & \sin{\bigl(\frac{\pi}{2}-\delta_{GW}\bigr)} \\
0 & 1 & 0\\
-\sin{\bigl(\frac{\pi}{2}-\delta_{GW}\bigr)} & 0 & \cos{\bigl(\frac{\pi}{2}-\delta_{GW}\bigr)}
\end{pmatrix} \\ \begin{pmatrix}
\cos{\omega_{e}\bigl(\alpha_{GW}-l_0-(t-t_0)\bigr)} & -\sin{\omega_{e}\bigl(\alpha_{GW}-l_0-(t-t_0)\bigr)} & 0\\
\sin{\omega_{e}\bigl(\alpha_{GW}-l_0-(t-t_0)\bigr)} & \cos{\omega_{e}\bigl(\alpha_{GW}-l_0-(t-t_0)\bigr)} & 0\\
0 & 0 & 1
\end{pmatrix} \\ \begin{pmatrix}
\cos{\bigl(\frac{\pi}{2}-\theta_{lat}\bigr)} & 0 & -\sin{\bigl(\frac{\pi}{2}-\theta_{lat}\bigr)} \\
0 & 1 & 0\\
\sin{\bigl(\frac{\pi}{2}-\theta_{lat}\bigr)} & 0 & \cos{\bigl(\frac{\pi}{2}-\theta_{lat}\bigr)}
\end{pmatrix}\begin{pmatrix}
\cos{\psi_0} & \sin{\psi_0} & 0\\
-\sin{\psi_0} & \cos{\psi_0} & 0\\
0 & 0 & 1
\end{pmatrix} .
\end{multline}

$\omega_e$ is the angular speed of Earth's rotation. $\alpha_{GW}$, $\delta_{GW}$ are the RA and DEC of the GW propagation direction. $\psi_{eq}$ is the polarization of the GW in the equatorial celestial coordinates. $\psi_0$ is the angle between the line joining the center of the circular accelerator to the timing detector and the longitude passing through the center of the circular accelerator. $l_0$ and $\theta_{lat}$ are respectively, the local sidereal time at the beginning of the observing run and latitude of the center of the circular accelerator.


%

\end{document}